\documentclass[conference]{IEEEtran}
\usepackage{cite}

\usepackage[cmex10]{amsmath}
\usepackage{array}
\usepackage{stmaryrd} 

\usepackage{mdwmath}
\usepackage{mdwtab}

\usepackage[footnotesize]{subfigure}
\usepackage[font=footnotesize]{subfig}

\usepackage{url}
\usepackage[dvips]{graphicx}
\usepackage{latexsym}
\usepackage{algorithm}
\usepackage{algorithmic}
\usepackage[smartEllipses]{markdown}
\usepackage{xcolor}
\usepackage{amsmath,amssymb,amsfonts}
\usepackage{amsthm}
\usepackage{multirow}
\usepackage{dirtree} 
\usepackage{listings}

\begin{document}

\definecolor{verylightgray}{rgb}{.97,.97,.97}

\lstdefinelanguage{Solidity}{
	keywords=[1]{anonymous, assembly, assert, balance, break, call, callcode, case, catch, class, constant, continue, constructor, contract, debugger, default, delegatecall, delete, do, else, emit, event, experimental, export, external, false, finally, for, function, gas, if, implements, import, in, indexed, instanceof, interface, internal, is, length, library, log0, log1, log2, log3, log4, memory, modifier, new, payable, pragma, private, protected, public, pure, push, require, return, returns, revert, selfdestruct, send, solidity, storage, struct, suicide, super, switch, then, this, throw, transfer, true, try, typeof, using, value, view, while, with, addmod, ecrecover, keccak256, mulmod, ripemd160, sha256, sha3}, 
	keywordstyle=[1]\color{blue}\bfseries,
	keywords=[2]{address, bool, byte, bytes, bytes1, bytes2, bytes3, bytes4, bytes5, bytes6, bytes7, bytes8, bytes9, bytes10, bytes11, bytes12, bytes13, bytes14, bytes15, bytes16, bytes17, bytes18, bytes19, bytes20, bytes21, bytes22, bytes23, bytes24, bytes25, bytes26, bytes27, bytes28, bytes29, bytes30, bytes31, bytes32, enum, int, int8, int16, int24, int32, int40, int48, int56, int64, int72, int80, int88, int96, int104, int112, int120, int128, int136, int144, int152, int160, int168, int176, int184, int192, int200, int208, int216, int224, int232, int240, int248, int256, mapping, string, uint, uint8, uint16, uint24, uint32, uint40, uint48, uint56, uint64, uint72, uint80, uint88, uint96, uint104, uint112, uint120, uint128, uint136, uint144, uint152, uint160, uint168, uint176, uint184, uint192, uint200, uint208, uint216, uint224, uint232, uint240, uint248, uint256, var, void, ether, finney, szabo, wei, days, hours, minutes, seconds, weeks, years},	
	keywordstyle=[2]\color{teal}\bfseries,
	keywords=[3]{block, blockhash, coinbase, difficulty, gaslimit, number, timestamp, msg, data, gas, sender, sig, value, now, tx, gasprice, origin},	
	keywordstyle=[3]\color{violet}\bfseries,
	identifierstyle=\color{black},
	sensitive=false,
	comment=[l]{//},
	morecomment=[s]{/*}{*/},
	commentstyle=\color{gray}\ttfamily,
	stringstyle=\color{red}\ttfamily,
	morestring=[b]',
	morestring=[b]"
}

\lstset{
	language=Solidity,
	backgroundcolor=\color{verylightgray},
	extendedchars=true,
	basicstyle=\footnotesize\ttfamily,
	showstringspaces=false,
	showspaces=false,
	numbers=left,
	numberstyle=\footnotesize,
	numbersep=9pt,
	tabsize=2,
	breaklines=true,
	showtabs=false,
	captionpos=b
	}

%
\title{Eth2Vec: Learning Contract-Wide Code Representations for Vulnerability Detection on Ethereum Smart Contracts}

\author{\IEEEauthorblockN{Nami Ashizawa}
\IEEEauthorblockA{Osaka University}
\and
\IEEEauthorblockN{Naoto Yanai}
\IEEEauthorblockA{Osaka University}
\and
\IEEEauthorblockN{Jason Paul Cruz}
\IEEEauthorblockA{Osaka University}
\and 
\IEEEauthorblockN{Shingo Okamura}
\IEEEauthorblockA{National Institute of Technology, Nara College}
}


%


\IEEEoverridecommandlockouts

\maketitle

\begin{abstract}
Ethereum smart contracts are programs that run on the Ethereum blockchain, and many smart contract vulnerabilities have been discovered in the past decade. Many security analysis tools have been created to detect such vulnerabilities, but their performance decreases drastically when codes to be analyzed are being rewritten. In this paper, we propose Eth2Vec, a machine-learning-based static analysis tool for vulnerability detection, with robustness against code rewrites in smart contracts. Existing machine-learning-based static analysis tools for vulnerability detection need features, which analysts create manually, as inputs. In contrast, Eth2Vec automatically learns features of vulnerable Ethereum Virtual Machine (EVM) bytecodes with tacit knowledge through a neural network for natural language processing. Therefore, Eth2Vec can detect vulnerabilities in smart contracts by comparing the code similarity between target EVM bytecodes and the EVM bytecodes it already learned. We conducted experiments with existing open databases, such as Etherscan, and our results show that Eth2Vec outperforms the existing work in terms of well-known metrics, i.e., precision, recall, and F1-score. Moreover, Eth2Vec can detect vulnerabilities even in rewritten codes. 
\end{abstract}


%

\section{Introduction} 

\subsection{Backgrounds} \label{Introduction}

Ethereum~\cite{theyellowpaper} is the largest platform that provides an execution environment for smart contracts, and many distributed applications have been developed and deployed as smart contracts on the Ethereum blockchain. 
Ethereum smart contracts are programs that are stored on the Ethereum blockchain and are run by the Ethereum Virtual Machine (EVM) as EVM bytecodes\footnote{Hereafter, ``Ethereum smart contract/s'' and ``smart contract/s'' are used interchangeably but have the same meaning.}. 

Given the transparent and decentralized nature of the Ethereum blockchain, the EVM bytecodes of smart contracts deployed on the Ethereum blockchain can be accessed and analyzed by anyone. 
Unfortunately, this also means that an adversary can abuse smart contracts~\cite{zou2019smart} by analyzing their EVM bytecodes and looking for vulnerabilities.
Consequently, attacks on vulnerable smart contracts can occur and possibly cause significant damage, especially when the attacked smart contracts handle assets. 
For example, the DAO attack is an infamous springboard attack where the attacker/s exploited a vulnerability in the DAO smart contract and stole more than 60 million USD worth of Ether, the cryptocurrency used in Ethereum.

According to literature~\cite{zou2019smart}, the security of smart contracts cannot be guaranteed because of the complexity (or lack of complexity) of the programming languages used for creating smart contracts, e.g., Solidity, which are relatively new languages, and the insufficient knowledge of programmers when creating smart contracts. 
To make matters worse, smart contracts that are successfully deployed on a blockchain cannot be modified, i.e., their source codes cannot be edited or deleted. In the past years, many attacks on deployed smart contracts have been reported~\cite{atzei2017survey}, and thus making sure that the source codes of smart contracts are not vulnerable before they are deployed on a blockchain is desirable. 
To do this, many analysis tools for vulnerability detection in Ethereum smart contracts have been developed~\cite{di2019survey}.

In this paper, we aim to develop a static analysis tool that can precisely identify various vulnerabilities in smart contract codes with high throughput by analyzing these codes. 
In static analysis, only a source code of a target to be analyzed is provided as input to identify if the code has vulnerabilities without executing the target itself.
Therefore, ideally, static analysis can be used to prevent vulnerable codes from being deployed.

However, static analysis has two problems in general: (1) accuracy of its vulnerability detection is limited, and (2) its analysis time can be long. 
For instance, disassembly from EVM bytecodes~\cite{norvill2018visual,zhou2018erays,suiche2017porosity,brent2018vandal} does not have the capability to identify whether a program is vulnerable or not, and the early literature focuses on simply improving the readability of disassembled codes. 
In other words, disassembled codes often need to be analyzed manually, consequently increasing the number of false positives and false negatives.
Moreover, symbolic execution~\cite{luu2016making,torres2018osiris,Weiss2019annotary,chinen2020hunting}, which extracts control flow graphs (CFGs) from a target code, achieves high accuacy by automating the analysis, but the generation of CFGs needs to search all states such that the target code transits. 
Therefore, the analysis takes significant amounts of time~\cite{ContractWard}. 

A potential solution to the problems described above is machine learning. Static analysis based on machine learning infers whether a given code is vulnerable by learning features of codes. 
In doing so, the analysis also achieves a versatile analysis within a short time.
However, static analysis based on machine learning has two inherent limitations: (1) code rewrites decrease analysis accuracy, and (2) appropriate features of smart contracts are indefinite. 
In the first limitation, for instance, CFGs with inlined functions are different from those of the original functions. Therefore, identifying the original function with the inlined function as the same codes is challenging. A pair of functions with the same semantics but different structures may generate different analysis results because the differences between structures of the codes strongly affect the analysis. 
In the second limitation, although features are manually extracted for machine learning, code samples and open knowledge about smart contracts are insufficient. 
Common knowledge about smart contract features has never been established~\cite{zou2019smart}. 
Notably, kinds of features representing vulnerabilities in codes of smart contracts are unknown. 
Besides, features that are robust against differences in code structures are still not obvious.

\subsection{Contributions}

In this paper, we propose \textit{Eth2Vec}, a static analysis tool based on machine learning that identifies smart contract vulnerabilities by learning smart contract codes via their EVM bytecodes, assembly codes, and abstract syntax trees. 
Eth2Vec has high throughput, high accuracy, and robustness against code rewrites.
Eth2Vec is an analysis tool based on a neural network for natural language processing, and it outputs the existence and kind of vulnerabilities in a target smart contract code only by taking the code as input. 
Using Eth2Vec, a user can analyze codes of smart contracts quickly even without expert knowledge on smart contract vulnerabilities.
To achieve this, we also developed a parser for EVM bytecodes, 
including compilation of the Solidity language, which is a high-level language used for creating smart contracts. 
As a result, developers can analyze their smart contract codes directly even without expert knowledge of vulnerabilities and before deploying them onto the blockchain.

In terms of analysis of Ethereum smart contracts by using machine learning, a major contribution of this paper is the provision of a method that is robust against code rewrites. 
As described in the previous subsection, analysis via a typical machine learning algorithm~\cite{momeni2019machine} may output wrong results when codes to be analyzed are rewritten. 
Such situation happens because existing tools learn \textit{only} patterns of code descriptions, i.e., the tools cannot learn the underlying features of the codes themselves. 
As another aspect of the limitation above, features that should be leveraged for analysis have never been established in existing works, to the best of our knowledge. 

Eth2Vec overcomes the limitation described above by leveraging a neural network for natural language processing. 
While existing tools~\cite{ContractWard,momeni2019machine} learn features that are given manually, Eth2Vec automatically learns features by entrusting the feature extraction phase to a neural network.
In other words, using a neural network can isolate the feature extraction from the technical difficulty of analysis of Ethereum smart contracts.
Indeed, a neural network for natural language processing has achieved high accuracy in processing of assembly codes~\cite{zuo2019neural,asm2vec}.
Nonetheless, Eth2Vec is novel for utilizing a neural network for the analysis of Ethereum smart contracts.
To do this, we also developed a module that gives the EVM bytecodes to the neural network as inputs. 
Furthermore, we designed a learning methodology based on a neural network for natural language processing which takes the EVM bytecodes through compilation of Solidity codes as inputs. 
Eth2Vec achieves robust analysis against code rewrites as well.

We conducted experiments to evaluate the performance of Eth2Vec. 
We used 5,000 files from Etherscan~\footnote{\url{https://etherscan.io}} as the dataset of contracts in Eth2Vec, and then executed the 10-fold cross-validation. 
The experimental results show that Eth2Vec can detect vulnerabilities within 1.2 seconds per contract with an average precision of 77.0\%. 
Notably, reentrancy, whose severity is the highest among known vulnerabilities~\cite{tikhomirov2018smartcheck}, can be detected with 86.6\% precision. 
Our results also indicate that Eth2Vec outperforms the method by Momeni et al.~\cite{momeni2019machine}, i.e., the use of support vector machine (SVM) based on their recommended features, as a naive method in terms of precision, recall, and F1-score of vulnerability detection. 
Besides, when we checked outputs by Eth2Vec in detail, we found examples of code clones with code rewrites and their vulnerabilities, which were not found by the SVM-based method in the current experiment. 


We plan to release the source codes of Eth2Vec via GitHub for reproducibility and as reference for future works. 

\section{Preliminaries} 
\label{preliminaries}

In this section, we describe background knowledge to help readers understand our work. 

\subsection{Ethereum Smart Contracts} 


In Ethereum, there are two kinds of accounts, namely, an externally owned account (EOA) and a contract account. EOAs have a private key that can be used to access the corresponding Ether or contracts. A contract account has smart contract code, which an EOA cannot have, and it does not have a private key. Instead, it is owned and controlled by the logic of its smart contract code.
In Ethereum, a smart contract refers to an \emph{immutable} computer program that is deployed on the blockchain and runs \emph{deterministically} in the context of the EVM. The immutability property indicates that, similar to any data published on a general blockchain, smart contract codes can be considered as trustworthy, i.e., once deployed, they cannot be changed or deleted. The deterministic property indicates that the execution of the coded functions of smart contracts will produce the same result for anyone who runs them.
Once deployed on the blockchain, a contract is self-enforcing and managed by the peers in the network, i.e., its functions are executed when the conditions in the contract are met. A smart contract is given an identity in terms of a contract address. Using this address, it can receive Ether and its functions can be executed. A contract is invoked when its contract address is the destination of a transaction, which is a signed message originating from an EOA, transmitted by the network, and recorded on the blockchain. Such transaction causes a contract to run in the EVM using the transaction (and transaction’s data) as input. The data indicate which specific function in the contract to run and what parameters to pass to that function. 
To incentivize peers to execute contract functions, Ethereum relies on \emph{gas}, which is paid in Ether, to “fuel computations”. The amount of gas needed to execute a transaction is relative to the complexity of the computations, thus also preventing infinite loops. 

Smart contracts are typically written in a high-level language, such as Solidity~\cite{solidityDoc}. 
The source code is then compiled to low-level bytecode that runs in the EVM. 
The EVM is a simple stack-based architecture. Its instruction set is kept minimal to avoid incorrect implementations that could cause consensus problems.
The EVM is a global singleton, i.e., it operates like a global, single-instance computer that runs in all peers in the network. Each peer runs a local copy of the EVM to validate the execution of contract functions, and the processed transactions and smart contracts are recorded on the blockchain.

\subsection{Machine Learning}  \label{machine learning}
Machine learning consists of two algorithms, i.e., \textit{training} and \textit{inference}. 
The training algorithm takes data as input to learn their features and optimize parameters inside the model for an objective function. 
On the other hand, the inference algorithm takes unseen data as input and infers a similar set of features to that of the training data. 
When each data is unlabeled, a learning algorithm is called unsupervised learning. 
The most popular approach to machine learning in the recent years is deep learning, which is based on neural networks and can extract features in a black-box manner. 
In this paper, we aim to develop a model for learning vulnerable smart contracts to detect vulnerabilities in unlearned smart contracts.

\section{Analysis of Smart Contracts \\ via Machine Learning} 
\label{problem description}

This section describes the analysis target as the problem setting and its technical difficulty to be tackled in this paper.

\subsection{Analysis Target} 

In this paper, we focus on security analysis of the Solidity language as a target of static analysis of Ethereum smart contracts. 
In particular, we aim to identify the existence of vulnerabilities and classify the kinds of vulnerabilities in the codes to be analyzed. 
This means that, for example, a developer uses a tool to analyze the smart contracts he/she is developing in local. 
Such tool potentially needs to convince a developer that a smart contract being developed does not have any vulnerability even if the developer does not have sufficient knowledge about smart contract vulnerabilities.
Therefore, 
we aim to develop a tool that can identify the existence of vulnerabilities (if there are any) in smart contracts even if 
only the codes of the smart contracts are given as input. 
Such a specification for analyzing smart contracts is preferable because standardized knowledge about Ethereum smart contracts is insufficient compared to general programming languages such as C and Java~\cite{zou2019smart}. 
Meanwhile, strong obfuscation, i.e., the use of encryption, on Solidity codes is out of the scope of this paper because such an obfuscated code of smart contracts does not exist as far as we know. 


Hereafter, we refer to codes written in Solidity as a contract and a file of codes consisting of more than a single contract as a contract file. 
We also call contracts to evaluate a vulnerability as test contracts and those to learn the vulnerability as training contracts. We call codes obtained from the compilation of the contracts as EVM bytecodes. 
The largest unit in each contract is a function, and a library function is also identical to a function. 
Finally, ``the blockchain'' will be used to refer to the Ethereum blockchain unless otherwise specified. 
\color{black}
The problem setting in this paper is then formalized as follows: 

\textbf{Problem Formulation:}
We formalize our approach for analysis of smart contracts as follows.
Each contract $c_i \in \mathcal{C}$ includes vulnerabilities $V_i =\{ v_1^i, \cdots, v_l^i\} \in \mathcal{V}^l$, 
where $\mathcal{C}$ denotes a set of contracts, $\mathcal{V}$ denotes a set of vulnerabilities independent of each other, and $l$ denotes any number.
Given any integer $n\in \mathbb{N}$,
a combination of a contract and vulnerabilities $CV= \{ (c_1, V_1), \cdots, (c_n, V_n)\}$ and a test contract $c_t \in \mathcal{C}$ are inputs of a model $M$.
Let $\{ \epsilon_i^{c_t} \}_{i \in [1,d]}  \subseteq \mathbb{R}^d$ denote the output of the model $M$ which has $d = |\mathcal{V}|$ elements, where $|\mathcal{V}|$ denotes the size of $\mathcal{V}$ and $\epsilon_i^{c_t}$ denotes a probability about vulnerabilities in $\mathcal{V}$.
Our goal is to develop a tool that optimizes $M (CV, c_t) \rightarrow\{ \epsilon_i^{c_t} \}$. 
\color{black}


\subsection{Technical Difficulty} \label{technical problem}

Although several machine-learning-based static analysis tools for smart contracts have been proposed in literature~\cite{ContractWard,momeni2019machine,liu2018eclone,liu2019enabling,song2019efficient}, accuracy of their vulnerability detection is limited even on known vulnerabilities.
In other words, accuracy significantly decreases when codes are rewritten from the original codes. 
Namely, on the problem formulation described above, the limitation described in Section~\ref{Introduction} as our motivation is formalized as, 
even if $c_t$ with a vulnerability $V_i$ which is included in $CV$ for any $i\in [1,n]$, $\epsilon_{i}^{c_{t}} \ll \epsilon_{i}^{c_{i}}$ holds for $c_t \notin CV$. 
We call such a situation non-\textit{robust}.



The limitation related to the accuracy of vulnerability detection described above is caused by insufficient extraction of features of the existing tools. 
In general, a machine learning model needs features as inputs, which are manually extracted, and then learns the features explicitly. 
However, features for representing smart contracts to be analyzed are non-obvious because the history of Ethereum smart contracts is shorter than other general languages such as C and Java~\cite{zou2019smart}. 
Intuitively, the limitation about the non-robustness is denoted as that required of a model $M$ is unknown. 
Another reason for the insufficient extraction of features is the lack of code samples of smart contracts~\cite{zou2019smart}.

\section{Design of Eth2Vec}
\label{proposed method}


In this section, we present Eth2Vec. 
We first describe the design concept to overcome the technical difficulty described in the previous section and then present the tool overview and its building blocks. 
Finally, we present the Eth2Vec model with its objective function.

\subsection{Design Concept} 

We aim to solve the technical difficulty by leveraging neural networks for natural language processing. 
Loosely speaking, neural networks handle feature extraction in a black-box manner, and thus the extraction of features can be isolated from the technical difficulty of the training. 
Therefore, Eth2Vec can analyze vulnerabilities of codes even if essential features of the vulnerabilities are unclear.

The term natural language processing described above means that the computation of the code similarity so that each word and paragraph are vectorized by inputting text data to neural networks, e.g., Word2Vec. 
When the security of codes is analyzed, a model learns vulnerable codes and it can then identify the vulnerabilities via the similarity of codes to be analyzed with the learned codes. 

To incorporate the natural language processing, 
Eth2Vec utilizes the \textit{PV-DM} model~\cite{le2014distributed} as neural networks to deal with EVM bytecodes. 
PV-DM model learns document representation based on tokens in the document. 
However, according to Ding et al.~\cite{asm2vec}, 
a document is sequentially laid out, which is different from program codes. 
In particular, program codes can be represented as a graph and has a specific syntax. 
On the design of Eth2Vec, toward the analysis of EVM bytecodes leveraging the PV-DM model, 
we developed a new module named \textit{EVM Extractor} to represent an abstract syntax tree of the codes. 

\subsection{Tool Overview} \label{Overview}

\begin{figure*}[h]
  \centering
  \includegraphics[width=\textwidth]{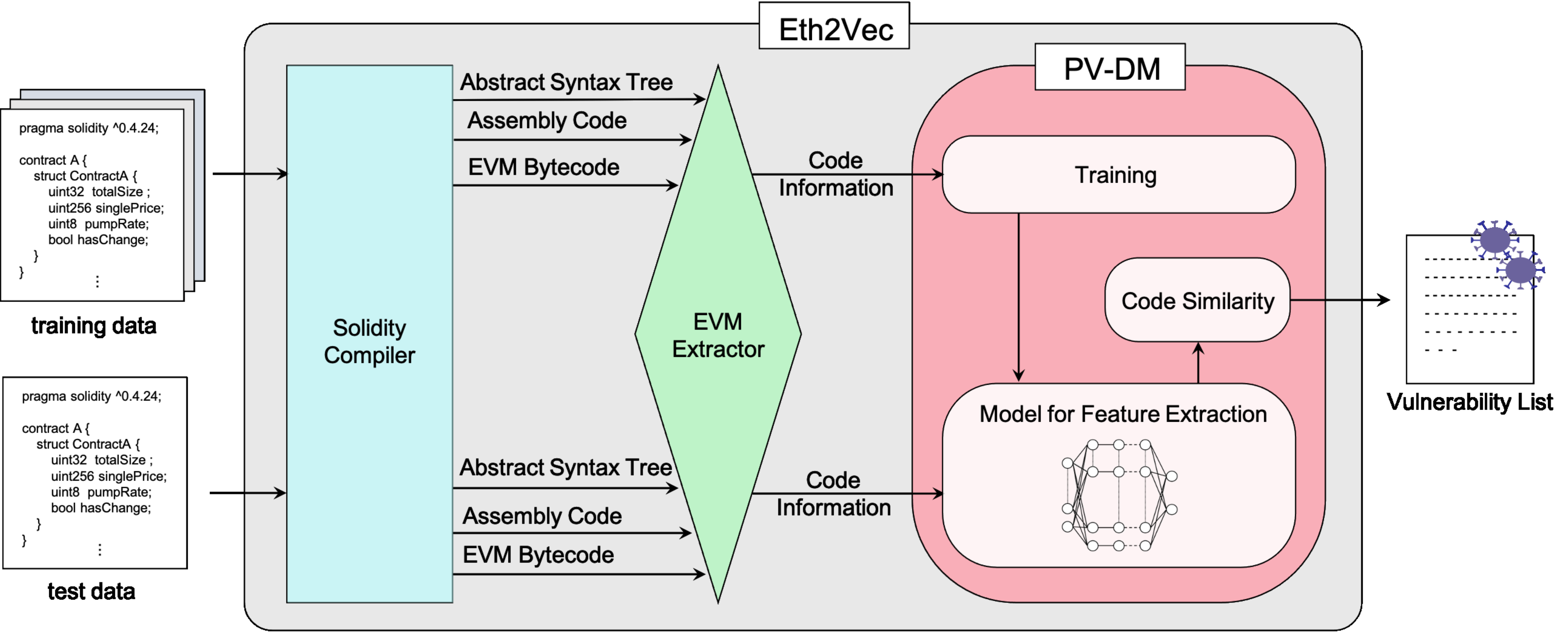}
  \caption{Overview of Eth2Vec}
  \label{fig:overview}
\end{figure*}

Eth2Vec consists of two modules, 
i.e., PV-DM model~\cite{le2014distributed} for neural networks to deal with paragraphs and EVM Extractor to create inputs of the PV-DM model from Solidity source codes. 

The overview of Eth2Vec is shown in Figure~\ref{fig:overview}.
First, a PV-DM model is utilized as neural networks to deal with bytecodes. 
In particular, the PV-DM model executes unsupervised learning by taking JSON files generated from bytecodes as input and then computes the code similarity for each contract. 
To do this, we developed the EVM Extractor as a module to create inputs of the PV-DM model from EVM bytecodes because the PV-DM model cannot deal with EVM bytecodes initially. 
More specifically, the EVM Extractor analyzes EVM bytecodes syntactically and creates JSON files for instruction-level, block-level, function-level, and contract-level. 

As a result, Eth2Vec takes EVM bytecodes to be analyzed as input, and then returns lists of code clones and their vulnerabilities for contract-level from a user's standpoint. 
Meanwhile, vulnerabilities of contracts for training data are identified in advance by the use of existing tools~\cite{luu2016making,tikhomirov2018smartcheck}. 
Vulnerabilities for test data are then evaluated by the code similarity with the vulnerable contracts. 
We show an output example of Eth2Vec in Appendix~\ref{output example}. 

\subsection{Building Blocks}

Eth2Vec utilizes the PV-DM model~\cite{le2014distributed} and EVM Extractor as building blocks. 

\subsubsection{PV-DM model} \label{PV-DM}

The PV-DM model is an extension of Word2Vec that treats text data for paragraph-level. 
Intuitively, it learns vector representations for each word and each paragraph. 
More concretely, given a text paragraph consisting of multiple sentences, a PV-DM model applies a sliding window over each sentence. 
The sliding window starts from the beginning of the sentence and moves forward a single word at each step. 
In doing so, the PV-DM model executes a multi-class prediction task such that it maps the current paragraph into a vector and each word in the context into a vector. 
More precisely, the model averages these vectors and infers the target word from the vocabulary via the softmax function. 
Formally, given a text that contains a list of paragraphs $p \in T$, each paragraph $p$ contains a list of sentences $ s \in p$, and each sentence is a sequence of $|s|$ words $w_t \in s$. 
Then, the PV-DM model maximizes the log probability as follows: 
\begin{equation} 
    \sum_{p}^{T} \sum_{s}^{p} \sum_{t =k}^{|s| -k} 
    \log {\textbf{P} (w_t |p, w_{t-k}, \cdots, w_{t+k})}    .  \label{objective of PV-DM}
\end{equation}
PV-DM model is initially designed for text data that is sequentially laid out while it can also be for analysis on an assembly code by extending vector representation of language~\cite{asm2vec}. 
In this paper, we present a contract-wide representation learning model by leveraging the PD-DM model on the syntax of EVM bytecodes.  

\subsubsection{EVM Extractor} \label{EVM Extractor}

EVM Extractor is a module that syntactically analyzes EVM bytecodes for PV-DM model. 
In particular, EVM Extractor parses Solidity files as in the following hierarchical structure: 
\dirtree{%
.1 data: dictionary. 
.2 name: file name.
.2 md5: md5 hash.
.2 functions: list.
.3 name: function name.
.3 sea: start point.
.3 see: end point.
.3 id: function id.
.3 call: list of callee functions.
.3 blocks: list.
.4 name: block name.
.4 bytes: bytecodes.
.4 sea: start point.
.4 eea: end point.
.4 id: block id.
.4 call: list of callee blocks.
.4 src: assembly instructions.
}
The top-level, \texttt{data}, represents a contract file to be analyzed. 
The second level indicates file-dependent information. The third level indicates function-dependent information.
The fourth level indicates a block consisting of multiple instructions. 
In each level, \texttt{sea} means the start address of the current function and that of the current block. 
Likewise, \texttt{see} and \texttt{eea} mean the start address of the next function and that of the next block, respectively. 
In contrast, \texttt{call} represents a callee function (or a callee block) from the current function (or block). 
Finally, at the bottom level, instructions of EVM bytecodes are stored in \texttt{src} together with their addresses. 
Meanwhile, a library function is treated as an individual contract.

\subsection{Design of Objective Function}

Eth2Vec is a model that learns and infers codes of Ethereum smart contracts through unsupervised learning. 
Intuitively, it can be regarded as an extension of Word2Vec with a wider vector representation of codes given its ability to target Ethereum smart contracts.

The objective function of Eth2Vec is designed by extending that of the original PV-DM model described in Equation~(\ref{objective of PV-DM}) along with the syntax shown in Section~\ref{EVM Extractor} in the same argument in~\cite{asm2vec}, which targets on an assembly language. 
Given a file of an Ethereum smart contract to be analyzed, Eth2Vec vectorizes the codes for instruction-level and then computes the code similarity. 
The process above is executed for block-level related to the instructions, function-level, and contract-level recursively, and therefore Eth2Vec can identify codes, whose distribution is similar to that of the training data, as a clone. 
Formally, an objective function is defined as follows: 
\begin{equation} \label{objective of eth2vec}
    \sum_{C_i}^{\texttt{Dict}} 
    \sum_{f_s}^{\mathcal{U}(C_i)}
    \sum_{seq_i}^{\mathcal{S}(f_s)} 
    \sum_{in_j}^{\mathcal{I}(seq_i)} \sum_{t_c}^{\mathcal{T}(in_j)} 
    \log {\textbf{P} (t_c | C_i, in_{j-1}, in_{j+1})}    , 
\end{equation}
where we denote by $\texttt{Dict}$ a contract file, by $C_i$ a contract, by $f_s$ a function, by $seq_i$ a block consisting of plural instructions, by $in_j$ each instruction, and by $t_c$ a token with respect to the current instruction. 
Likewise, we denote by $\mathcal{U}(C_i)$ plural functions, 
by $\mathcal{S}(f_s)$ plural blocks, by $\mathcal{I}(seq_i)$ a list of instructions, and by $\mathcal{T}(in_j)$ a list of tokens.

Here, 
the first summation term, i.e., $\sum_{C_i}^{\texttt{Dict}}$, is included in the second level shown in the previous section, i.e., \texttt{file name}. 
Equation~(\ref{objective of eth2vec}) is given contracts and their instructions, and then it maximizes the log probability for the current token $t_c$. 
Intuitively, the lexical meaning for each contract is computed through the current instruction and its neighbor instructions. 
Moreover, codes consisting of plural contracts can also be analyzed by extracting features for each contract. 

We describe the objective function to represent the aforementioned intuition in detail below. 
Given a contract file $Dict$, a function $f_s$ for each contract $C_i$ is vectorized and then we denote by $\overrightarrow{\theta_{f_s}}$ the vector representation of $f_s$. 
Furthermore, we denote by $\mathcal{CT}(in)$ average of the vector representation of neighbor instructions of $in$. 
We then define a concatenation of the vector representation of the instruction itself and that of its operand as follows: 
\begin{equation} \label{vector-representation}
\mathcal{CT}(in)=\overrightarrow{v}_{\mathcal{P}(in)} || \frac{1}{\mathcal{A}(in)} \sum_t^{\mathcal{A}(in)} \overrightarrow{v_{t}}, 
\end{equation}
where $\mathcal{P}(in)$ is one operation with respect to $in$, 
$\mathcal{A}(in)$ is a list of operands with respect to $in$, and $||$ is a concatenation of strings. 
In doing so, for any contract $C_i$, by averaging $f_s$ and the $j$-th instruction $in_j$ with $\mathcal{CT}(in_{j-1})$ and $\mathcal{CT}(in_{j+1})$, 
a function $\delta (in_j, f_s)$ to evaluate the joint memory of neighbor instructions in $f_s$ is defined as follows: 
\begin{equation} \label{delta_function}
\delta (in_j, f_s)=\frac{1}{3}\left( \overrightarrow{\theta_{f_s}} + \mathcal{CT}(in_{j-1}) + \mathcal{CT}(in_{j+1}) \right).  
\end{equation}
Therefore, the log probability in Equation~(\ref{objective of eth2vec}) can be replaced with $\textbf{P} (t_c | C_i, in_{j-1}, in_{j+1}) = \textbf{P} (t_c | \delta(in_{j}, f_s) )$. 
According to the literature~\cite{asm2vec}, by letting $X=(\overrightarrow{v_{t_c}})^T \times \delta (in_j, f_s)$ and utilizing a sigmoid function $\sigma (X)=\frac{1}{1+e^{-X}}$, 
the above computation can be approximated by the $k$-negative sampling~\cite{mikolov2013distributed,le2014distributed} as follows: 
\begin{eqnarray}
&&
\sum_{C_i}^{Dict} 
\sum_{f_s}^{\mathcal{U}(C_i)}
\sum_{seq_i}^{\mathcal{S}(f_s)} 
\sum_{in_j}^{\mathcal{I}(seq_i)} \sum_{t_c}^{\mathcal{T}(in_j)}
J(\theta) \nonumber \\ 
&=& 
\sum_{C_i}^{Dict} 
\sum_{f_s}^{\mathcal{U}(C_i)}
\sum_{seq_i}^{\mathcal{S}(f_s)} 
\sum_{in_j}^{\mathcal{I}(seq_i)} \sum_{t_c}^{\mathcal{T}(in_j)}\log{  \textbf{P} \left( t_c | \delta ( in_{j}, f_s ) \right) } ,
\end{eqnarray}
\begin{eqnarray}
J(\theta) 
\approx  \label{objective of training}
\log{ \left( \sigma \left( X  \right) \right)   }
+ \sum_{i=1}^k \mathbb{E}_{t_d \sim P_n (t_c)} 
\left( 
\llbracket t_d \neq t_c \rrbracket \log{  \sigma \left( X \right) } \right), 
\end{eqnarray}
where $\llbracket t_d \neq t_c \rrbracket$ is an identity function which returns $1$ if the expression inside the function is true or $0$ otherwise. 
On the other hand, $\mathbb{E}_{t_d \sim P_n (t_c)} $ is a sampling function that samples a token $t_d$ in accordance with the noise distribution $P_n (t_c)$ from $t_c$. 
We finally utilize Equation~(\ref{objective of training}) as the objective function of Eth2Vec. 
Intuitively, the function maximizes the probability of a token $t_c$ of the current instruction and decreases that of a token $t_d$ of the other instructions. 
Then, we can compute the gradient with respect to $\theta_{f_s}$ through the following derivative: 
\begin{eqnarray}
\frac{\partial J (\theta) }{\partial \overrightarrow{\theta_{f_s}}} 
&=& \frac{\partial J (\theta) }{\partial X} \frac{\partial X }{\partial \overrightarrow{\theta_{f_s}}} \nonumber \\
&=& \frac{\overrightarrow{v_{t_c}}}{3} 
\left( 1 - \sigma \left( X \right) \right) \nonumber \\ 
&+&
\frac{\overrightarrow{v_{t_c}}}{3} \sum_i^k \mathbb{E}_{t_d \sim P_n (t_c)} 
\left( 
\llbracket t_d \neq t_c \rrbracket \left( 1- \sigma \left( X  \right) \right)
\right) \label{gradient_f} \nonumber
\end{eqnarray}
Moreover, we can also utilize a more intuitive gradient by approximating the aforementioned gradient computation as follows: 
\begin{equation}
\frac{\partial J (\theta) }{\partial \overrightarrow{\theta_{ f_s}}}    \approx \frac{\overrightarrow{v_{t_c}}}{3} \sum_i^k \mathbb{E}_{t \sim P_n (t_c)} 
\left( 
\llbracket t = t_c \rrbracket  - \sigma \left( X  \right) 
\right) \label{gradient_f-approx}. 
\end{equation}
Intuitively, Equation~(\ref{gradient_f-approx}) approximates the original derivative described above in a manner that it moves a gradient to the positive if a token is identical to the current instruction or to the negative otherwise. 
Likewise, the gradient with respect to the vector representation $\overrightarrow{v}_{t_c}$ of a token on the current instruction can be approximated. 
Moreover, for the vector representation $\overrightarrow{v}_{\mathcal{P}(in)}$ of a token of some operation $\mathcal{P}(in)$ and that $\overrightarrow{v}_{\mathcal{A} (in)}$ of operands on the current instruction, their gradients can be approximated in a similar manner. 
These gradients are computed as follows although we omit their derivation: 
\begin{eqnarray}
\frac{\partial J (\theta) }{\partial \overrightarrow{v_{t_c}}} 
&\approx& 
\left( 
\llbracket t = t_c \rrbracket  - \sigma \left( X
\right) 
\right) \cdot \delta (in_j, f_s) \label{gradient_v-approx} \\
\frac{\partial J (\theta) }{\partial \overrightarrow{v_{\mathcal{P} (j+1)}}} 
&\approx& 
\left( \llbracket t = t_c \rrbracket  - \sigma \left( X \right) \right) \cdot \frac{1}{\mathcal{A}(in)} \sum_t^{\mathcal{A}(in)} \overrightarrow{v_{t}} \label{gradient_vP-approx} \\ 
\frac{\partial J (\theta) }{\partial \overrightarrow{v_{\mathcal{A}(in_{j+1})}}} 
&\approx& 
\left( \llbracket t = t_c \rrbracket  - \sigma \left( X \right) \right) \cdot \left(  
\overrightarrow{v_{\mathcal{P}(in_{j+1} )}} \right) \label{gradient_v_t_b} 
\end{eqnarray}
Similar equations hold for the previous instruction $in_{j-1}$ by replacing $in_{j+1}$ with $in_{j-1}$ although we omit the details due to space limitation. 

\subsection{Training and Inference}

We now present the training algorithm and the inference algorithm with the gradients described in the previous section. 
First, the training algorithm is shown in Algorithm~\ref{eth2vec:training}. 
The goal of the training is to optimize vectors for each instruction belonging to the given contract file as input. 
Instructions with similar meaning are mapped to a similar position in the vector space through the training. 
These resulting vectors by the algorithm can be used as features for vector representation of contract-level analysis. 
The algorithm is an unsupervised training algorithm, and thus the procedure does not require a ground-truth mapping between equivalent contracts.


The aforementioned features can be utilized directly in the inference algorithm shown in Algorithm~\ref{eth2vec:infer}. 
In particular, for an unseen contract $C_t \notin \texttt{Dict}$, the algorithm first initializes $\overrightarrow{\theta_{ f_t}}$, which is associated with any function $f_t$ belonging to $C_t$. 
Then, the algorithm follows the same procedure as the training algorithm. 
However, all $\overrightarrow{v_t}$'s in the trained model are kept, and $\overrightarrow{\theta_{ f_t}}$'s are updated following their errors. 
At the end of the inference, $\overrightarrow{\theta_{ f_t}}$ is output whereas the vectors for all $C_i \in \texttt{Dict}$ remain the same except for $\overrightarrow{\theta_{ f_t}}$. 
Finally, to search contracts for a match with the given contract $C_t$, 
the resultant vectors are compared with those of the training contracts using a typical statistical method, e.g., the cosine similarity.

\begin{algorithm}[t]
  \caption{Training algorithm \texttt{TRAIN} of Eth2Vec for one epoch}
  \label{eth2vec:training}
  \begin{algorithmic}[1]
    \REQUIRE \texttt{Dict}
    \ENSURE vector representations for tokens of any $in_j$ and $\overrightarrow{\theta_{ f_s}}$
    \FORALL{$C_i \in \texttt{Dict}$}
        \FORALL{$f_s \in \mathcal{U}(C_i)$}
            \FORALL{$seq_i \in \mathcal{S}(f_s)$}
                \FORALL{$in_j \in \mathcal{I}(seq_i)$}
                    \STATE vectorize $f_s$ as $\overrightarrow{\theta_{ f_s}}$
                    \STATE compute $\mathcal{CT}(in_{j+1})$ by Eq.~(\ref{vector-representation})
                    \STATE compute $\mathcal{CT}(in_{j-1})$ by Eq.~(\ref{vector-representation})
                    \STATE compute $\delta(in_j, f_s)$ by Eq.~(\ref{delta_function})
                    \FORALL{$t_c \in \mathcal{T}(in_j)$}
                        \STATE compute a gradient for $\overrightarrow{\theta_{ f_s}}$ by Eq.~(\ref{gradient_f-approx})
                        \STATE compute a gradient for $\overrightarrow{v_{t_c}}$ by Eq.~(\ref{gradient_v-approx})
                        \STATE compute a gradient for $\overrightarrow{v_{\mathcal{P} (j+1)}}$ by Eq.~(\ref{gradient_vP-approx})
                        \STATE compute a gradient for $\overrightarrow{v_{\mathcal{A}(in_{j+1})}}$ by Eq.~(\ref{gradient_v_t_b})
                        \STATE compute a gradient for $\overrightarrow{v_{\mathcal{P} (j-1)}}$ by Eq.~(\ref{gradient_vP-approx})
                        \STATE compute a gradient for $\overrightarrow{v_{\mathcal{A}(in_{j-1})}}$ by Eq.~(\ref{gradient_v_t_b})
                    \ENDFOR
                    \STATE update vectors for tokens of $in_j$
                    \STATE update $\overrightarrow{\theta_{ f_s}}$
                \ENDFOR
            \ENDFOR
        \ENDFOR
    \ENDFOR
  \end{algorithmic}
\end{algorithm}
\begin{algorithm}[t]
  \caption{Inference algorithm \texttt{Infer} of Eth2Vec for a query}
  \label{eth2vec:infer}
  \begin{algorithmic}[1]
    \REQUIRE contract $C_t \notin \texttt{Dict}$
    \ENSURE vector representation $\overrightarrow{\theta_{f_t}}$ for any $f_t \in \mathcal{U}(C_t)$
    \STATE Initialize vector representation $\overrightarrow{\theta_{f_t}}$ of $f_t \in \mathcal{U}(C_t)$
    \FORALL{$f_t \in \mathcal{U}(C_t)$}
        \FORALL{$seq_i \in \mathcal{S}(f_t)$}
            \FORALL{$in_j \in \mathcal{I}(seq_i)$}
                \STATE compute $\mathcal{CT}(in_{j+1})$ by Eq.~(\ref{vector-representation})
                \STATE compute $\mathcal{CT}(in_{j-1})$ by Eq.~(\ref{vector-representation})
                \STATE compute $\delta(in_j, f_s)$ by Eq.~(\ref{delta_function})
                \FORALL{$t_c \in \mathcal{T}(in_j)$}
                    \STATE compute a gradient for $\overrightarrow{\theta_{ f_t}}$ by Eq.~(\ref{gradient_f-approx})
                    \STATE update $\overrightarrow{\theta_{ f_t}}$
                \ENDFOR
                \ENDFOR
            \ENDFOR
        \ENDFOR
      \end{algorithmic}
\end{algorithm}

\subsection{Implementation} \label{implementation}

We implement Eth2Vec by utilizing \texttt{Kam1n0}\footnote{Kam1n0 version 2.0.0: \url{https://github.com/McGill-DMaS/Kam1n0-Community}} and \texttt{py-solc-x}\footnote{py-solc-x: \url{https://pypi.org/project/py-solc-x/}}. 
First, the main module, PV-DM model, is implemented by \texttt{Kam1n0}, 
which is a server system~\cite{Ding2016Kam1n0} utilized for on binary analysis~\cite{asm2vec}. 
We mainly modified the source codes in \texttt{DisassemblyFactoryIDA.java} and \texttt{ExtractBinaryViaIDA.py}. 
In particular, \texttt{ExtractBinaryViaIDA.py} initially generates a JSON file extracted from IDA by disassembling binary codes, and then \texttt{DisassemblyFactoryIDA.java} takes the file to store the binary codes within \texttt{Kam1n0}. 
However, IDA cannot use EVM bytecodes for implementing Eth2Vec. 
Therefore, we changed \texttt{ExtractBinaryViaIDA.py}: For instance, the code information is obtained by compiling a Solidity file with \texttt{py-solc-x} without IDA, and then its resultant assembly codes, abstract syntax tree (AST), and binary codes are extracted.  
We plan to release the source codes of the whole implementation publicly via GitHub. 

\section{Experiments} \label{experiments}

In this section, we describe the experiments we conducted to evaluate Eth2Vec. 
First, we describe the purpose of the experiments. Then, we discuss the datasets and the training methodologies used for evaluation. 
Finally, we show the experimental results. 

\subsection{Purpose of Experiments}

To evaluate the performance of Eth2Vec, we try to identify vulnerabilities in codes to be analyzed through the training with the known vulnerable contracts. 
To do this, we first check if Eth2Vec appropriately represents the relationship between codes to be analyzed and codes learned in the training phase. 
We evaluate clone detection of codes written in the Solidity language to confirm whether Eth2Vec can appropriately extract features of the codes. In doing so, we also evaluate semantic clones based on the lexical-semantic relationship to confirm the robustness of Eth2Vec against code rewrites. 


Next, we check whether all the learned vulnerabilities are identified by training the codes and their code similarity with a given code. 
In doing so, we also evaluate consistency of the vulnerability detection with the results of the clone detection described above. 
In particular, we check if an output of the vulnerability detection is identical to the vulnerabilities of code identified as clones and if the clones are contained in the clone detection's output.


We confirm that Eth2Vec can extract features precisely and thus can detect vulnerabilities through the experiments mentioned above. 
We also compare the performance of Eth2Vec with that of the existing work by Momeni et al.~\cite{momeni2019machine}, which extracts features manually, as a baseline.  


\subsection{Experimental Setting}


As mentioned above, experiments are conducted in two stages, i.e., clone detection and vulnerability detection of smart contracts. 
We describe the setting for each experiment below. 
We first describe datasets and the baseline in detail.

\subsubsection{Dataset}


We collect 5,000 contract files from Etherscan\footnote{\url{https://etherscan.io/}}, which is an open database of smart contracts, as a dataset utilized in the experiments. 
These 5,000 files are also identical to files utilized in a recent work~\cite{gao2020smartembed}. 
We then utilize only the files that can be compiled by solidity version -0.4.11 as a compiler. 
The dataset contains 95,152 contracts and 1,193,868 blocks. 

\subsubsection{Baseline} \label{baseline}


We compare the performance of Eth2Vec with that of the scheme based on support vector machine (SVM) by Momeni et al.~\cite{momeni2019machine} as a baseline. 
Although there are several versatile results based on machine learning~\cite{ContractWard,liu2018eclone,gao2020smartembed}, which can detect various vulnerabilities, their source codes are unpublished or we were unable to build their source codes in our environment. 
We thus adopt the SVM-based method by Momeni et al. which we were able to reproduce from scratch. 

Momeni et al. extracted 16 features from an abstract syntax tree (AST) of an Ethereum smart contract. 
We utilize 15 of these features excluding hexadecimal addresses because hexadecimal addresses cannot be obtained from source codes, i.e., without deployment. 
Other features are described in Appendix ~\ref{feature by Momeni}. 

\subsubsection{Clone Detection}

We check if Eth2Vec can precisely identify clones of test contracts as input through training contracts. 
In particular, on the 10-fold cross-validation, 500 test contracts are randomly chosen, and the remaining 4,500 contracts are utilized as the training contracts. This process is iterated ten times. 
Meanwhile, a threshold of the code similarity to detect clones is 0.8. 

In the setting mentioned above, precision is utilized as an evaluation metric. 
For clones for each function, i.e., function-level clones, which are output by a threshold more than 0.8, we check if a function in a contract, which has the highest similarity, is indeed a clone of the given input.
To compare the performance of Eth2Vec, we also evaluate the SVM-based method described above by manually labeling each function. 
Meanwhile, clones in this experiment correspond to the type-I to type-I\hspace{-.1em}V clones in literature~\cite{Roy07asurvey}, and semantic clones are identical to the type-I\hspace{-.1em}V clones.  



\subsubsection{Vulnerability Detection}

We check if Eth2Vec can precisely detect vulnerabilities of test contracts as input through learning the training contracts. 
In particular, we check whether vulnerabilities output by Eth2Vec with a threshold 0.8 about the test contracts are identical to true vulnerabilities of the contracts.  


In this experiment, we adopt the 10-fold cross-validation similar to that in the clone detection experiment. 
We also confirm the vulnerabilities of the test contracts as ground truth of the experiment by utilizing Oyente~\cite{luu2016making} and SmartCheck~\cite{tikhomirov2018smartcheck}. 
The vulnerabilities are listed in Appendix~\ref{vulnerability list} due to space limitation. 
In this experiment, we adopt well-known metrics, i.e., accuracy, precision, recall, and F1-score.

\subsection{Results on Clone Detection} \label{results on clone detection}

The results of the clone detection by Eth2Vec are shown in Table~\ref{tab:clone}. 
\begin{table}[t]
  \begin{center}
    \caption{Precision of Clone Detection of Eth2Vec. This table shows the average and standard deviation of 10 executions of precision measurement on 10-fold cross-validation. 
    Furthermore, the row of SVM w/o few clones represents result of the SVM-based setting computed by removing few clones. 
    The numbers are truncated to one decimal place. 
    }
    \label{tab:clone}
    \begin{tabular}{|l|c|c|c|}
      \hline
      & Eth2Vec & SVM~\cite{momeni2019machine} & SVM w/o few clones \\ \hline \hline
      Average & 74.9\% & 34.6\% & 42.7\% \\  \hline
      Standard Deviation & 0.9 & 34.6 & 43.6 \\ \hline
    \end{tabular}
  \end{center}
\end{table}
According to the table, Eth2Vec can implicitly extract more features than the SVM-based method~\cite{momeni2019machine}. 
For instance, the values of the standard deviation by Eth2Vec are significantly smaller than those by the SVM-based method. 
Intuitively, this means that feature extraction by Eth2Vec is more stable, and therefore Eth2Vec is able to represent features on various codes. 
When the clone detection performance is compared to  EClone~\cite{liu2018eclone} as a reference, the detection rate of EClone with a threshold of 0.4 was 58.2\%. 
Although we did not implement EClone by ourselves as described in Section~\ref{baseline}, we believe that Eth2Vec can detect clones of Ethereum smart contracts better than EClone because of its more sophisticated feature extraction. 


When we checked an output of Eth2Vec in detail, it contained semantic clones. 
We show a concrete example in Listing~\ref{originalTransfer} and Listing~\ref{cloneTransfer}. 
The output of Eth2Vec, i.e., Listing~\ref{cloneTransfer}, is a clone of the original function shown in Listing~\ref{originalTransfer} excluding lines~5 and~9. 
Listing~\ref{cloneTransfer} is identical to a type-I\hspace{-.1em}I\hspace{-.1em}I clone of Listing~\ref{originalTransfer} according to the definitions in~\cite{Roy07asurvey}. 
This means that Eth2Vec can detect clones precisely, even those of rewritten codes. 


\begin{figure}[tb]
    \begin{lstlisting}[frame=single,label=originalTransfer,caption=Original function of Transfer,language=Solidity]
function _transfer(address _from, address _to, uint _value) internal {
    require(_to != 0x0);
    require(balanceOf[_from] >= _value);
    require(balanceOf[_to] + _value > balanceOf[_to]);
    uint previousBalances = balanceOf[_from] + balanceOf[_to];
    balanceOf[_from] -= _value;
    balanceOf[_to] += _value;
    Transfer(_from, _to, _value);
    assert(balanceOf[_from] + balanceOf[_to] == previousBalances);
}
    \end{lstlisting}
    \begin{lstlisting}[frame=single,label=cloneTransfer,caption=Clone of the original function of Transfer found by Eth2Vec,language=Solidity]
function _transfer(address _from, address _to, uint _value) internal {
    require(_to != 0x0); 
    require(balanceOf[_from] >= _value);
    require(balanceOf[_to] + _value > balanceOf[_to]);
    balanceOf[_from] -= _value;
    balanceOf[_to] += _value;
    Transfer(_from, _to, _value);
}
    \end{lstlisting}
\end{figure}

Meanwhile, the precision of the SVM-based method is small because the test datasets contain few clones in several executions. 
The SVM-based method infers negatives for the given test data under such situation.  Therefore, the precision, the recall, and the F1-score become $0$, resulting in a low average and a more extensive standard deviation. 
Nonetheless, when we re-compute the precision by removing the results with the few clones among ground-truth, the precision is updated to 42.7\%. 


\subsection{Results on Vulnerability Detection}

\begin{table*}[htb]
  \begin{center}
    \caption{Vulnerability Detection Results on Eth2Vec: In the following table, we measured well-known metrics for both Eth2Vec and the SVM-based method~\cite{momeni2019machine}. The values of the averages and standard deviations are computed by the values of each row. 
    The numbers are truncated to one decimal place.
    }
    \label{tab:vul}
    \begin{tabular}{|c|c||c|c|c||c|c|c|}
      \hline
       & & \multicolumn{3}{c||}{Eth2Vec}  &  \multicolumn{3}{c|}{SVM~\cite{momeni2019machine}} \\ \cline{3-8}
     Vulnerability & Severity & Precision [\%] & Recall [\%] & F1-Score [\%] & Precision [\%] & Recall [\%] & F1-Score [\%] \\ \hline \hline
      Reentrancy & 3 & 
      86.6 & 54.8 & 61.5 & 
      30.0 & 7.8 & 12.3 \\ \hline
      Time Dependency & 2 & 
      75.2 & 17.0 & 27.3 & 
      55.0 & 2.8 & 5.3 \\ \hline
      ERC-20 Transfer & 1 & 
      95.6 & 58.4 & 72.4 & 
      89.0 & 95.3 & 92.0 \\ \hline
      Gas Consumption & 1 & 
      48.0 & 29.0 & 32.4 & 
      10.0 & 3.1 & 4.7 \\ \hline
      Implicit Visibility & 1 & 
      68.9 & 82.0 & 74.8 & 
      71.5 & 77.5 & 73.8 \\ \hline
      Integer Overflow & 1 & 
      89.9 & 57.6 & 70.1 & 
      84.9 & 73.1 & 78.3 \\ \hline
      Integer Underflow & 1 & 
      74.6 & 56.0 & 63.7 & 
      75.1 & 39.2 & 50.0 \\ \hline
      \hline
      \multicolumn{2}{|c||}{Average} & 
      77.0 & 50.7 & 57.5 & 
      59.3 & 42.7 & 45.2 \\ \hline
      \multicolumn{2}{|c||}{Standard Deviation} & 
      14.7 & 19.7 & 18.0 & 
      27.3 & 36.4 & 34.7 \\ \hline
    \end{tabular}
  \end{center}
\end{table*}

The results of the vulnerability detection by Eth2Vec are shown in Table~\ref{tab:vul}.

First, the standard deviations of Eth2Vec are lower than those of the SVM-based method. 
This means that Eth2Vec diminishes the effect of difference between features of vulnerabilities because Eth2Vec can detect the vulnerabilities more stably than the SVM-based method without 
Eth2Vec can precisely extract features and thus, it can take into account various vulnerabilities. 


Second, we confirm that Eth2Vec is also robust against the dispersion of data. 
In particular, the appearance of vulnerabilities with high severity, i.e., reentrancy and time dependency, is low in the current dataset. 
Consequently, the number of the corresponding labels is less, and the recall of the SVM-based method became low. 
By contrast, the recall of Eth2Vec is more stable than that of the SVM-based method even against vulnerabilities, whose frequency of appearance is low,
in comparison with the other vulnerabilities. 


Moreover, when we checked an output of Eth2Vec in detail, we found a concrete example, as shown in Listing~\ref{vulnerableTransfer}. 
The code is output as a clone of Listing~\ref{originalTransfer} and the code similarity of Listing~\ref{originalTransfer} with Listing~\ref{vulnerableTransfer} is lower than that with Listing~\ref{cloneTransfer}. 
As difference between these codes, Listing~\ref{originalTransfer} and Listing~\ref{cloneTransfer} do not have vulnerabilities while Listing~\ref{vulnerableTransfer} contains the integer overflow vulnerability. 
Specifically, line~4 in Listing~\ref{vulnerableTransfer} is different from Listing~\ref{originalTransfer} with respect to an expression statement, and the statement in Listing~\ref{vulnerableTransfer} is identical to a typical form of integer overflow\footnote{\url{https://github.com/ConsenSys/mythril/wiki/Integer-Overflow}}. 


The aforementioned example indicates that the vulnerability detection of Eth2Vec is robust against code rewrites. 
Although Listing~\ref{originalTransfer} and Listing~\ref{vulnerableTransfer} seem to be more similar than Listing~\ref{originalTransfer} and Listing~\ref{cloneTransfer}, 
Listing~\ref{vulnerableTransfer} is different from a ``precise" clone of Listing~\ref{originalTransfer} because Listing~\ref{vulnerableTransfer} contains the integer vulnerability, which is not included in Listing~\ref{originalTransfer}. 
Therefore, Eth2Vec returns a higher code similarity for Listing~\ref{cloneTransfer} and Listing~\ref{originalTransfer} than for Listing~\ref{vulnerableTransfer} and Listing~\ref{originalTransfer}. 
We thus confirm that Eth2Vec has robust vulnerability detection. 

\begin{figure}[tb]
    \begin{lstlisting}[frame=single,label=vulnerableTransfer,caption=Vulnerability example by Eth2Vec to the original function,language=Solidity]
function _transfer(address _from, address _to, uint _value) internal {
    require(_to != 0x0);
    require(balanceOf[_from] >= _value);
    require(balanceOf[_to] + _value >= balanceOf[_to]);
    uint previousBalances = balanceOf[_from] + balanceOf[_to];
    balanceOf[_from] -= _value;
    balanceOf[_to] += _value;
    emit Transfer(_from, _to, _value);
    assert(balanceOf[_from] + balanceOf[_to] == previousBalances);
}
    \end{lstlisting}
\end{figure}



\subsection{Throughput for Inference} \label{inference}

\begin{figure}[h]
  \centering
  \includegraphics[width=\columnwidth]{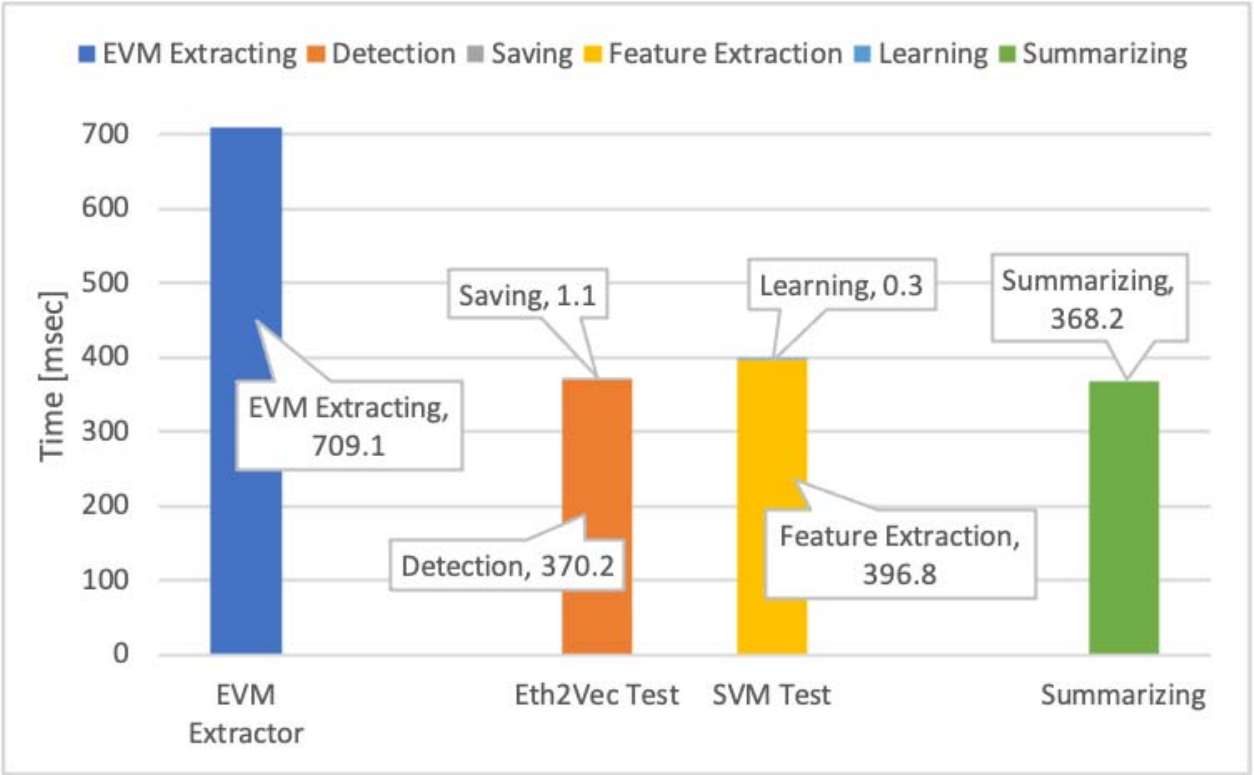}
  \caption{Throughput of Eth2Vec: We measured throughput of Eth2Vec for each process. The item of Eth2Vec Test is identical to the vulnerability detection, which contains the detection itself and saving detection results. 
  The item of Summarizing represents the processing time for the display process of the saved data. 
  The entire process until a user obtains an output is a summation of EVM Extractor, Eth2Vec Test, and Summarizing.
  On this measurement, we randomly picked up 20 contract files, where a threshold is 0.8 and five clones, i.e., candidates of vulnerabilities, at maximum for each function are output. 
    }
  \label{fig:throughput}
\end{figure}

The throughput of the inference by Eth2Vec is shown in Figure~\ref{fig:throughput}. 
The processing time for vulnerability detection depends on the number of functions and that for displaying the detection results, i.e., summarizing by \texttt{Kam1n0}, depends on the number of detected clones. 
According to the current measurement, the detection throughput of Eth2Vec is faster than that of the SVM-based method except for the process of EVM Extractor. 
Eth2Vec can analyze within about 0.371 seconds per contract in comparison with about 0.397 seconds per contract by the SVM-based method. 
Meanwhile, the processing time for displaying the detection results is longer than the detection itself.
A long analysis needed 5.6 seconds for detection, 0.015 seconds for saving, and 3.5 seconds for summarizing, where the number of contracts is nine and the number of functions is 94.

\subsection{Limitations}

In this section, we describe several limitations of Eth2Vec, which will be improved in future works. 
First, the current construction is in the unsupervised setting, and the supervised setting is not implemented. 
According to Hill et al.~\cite{hill2016learning}, the supervised setting seems to be more suitable for the classification of natural language processing than the unsupervised setting. 
Intuitively, evaluating a kind of vulnerability contained in codes is a classification problem about vulnerabilities. 
Namely, the performance of Eth2Vec can be improved by utilizing the supervised setting. 

Second, the current work only detected vulnerabilities of contracts included in the training dataset utilized in our experiment. 
There are potentially many vulnerable contracts that have already been deployed, and thus it would be ideal if all deployed contracts are analyzed to determine potential vulnerabilities.


Finally, the current construction does not support inter-contract analysis where multiple contracts are interconnected. 
For instance, Rodler et al.~\cite{rodler2018sereum} presented a new kind of vulnerability by utilizing CALL and CREATE instructions, which requires inter-contract analysis~\cite{Weiss2019annotary,chinen2020hunting}. Thus, the attacks by Rodler et al. are out of the scope of Eth2Vec currently. 
Our future work will aim to extend the current construction of Eth2Vec and overcome the limitations described above.

\section{Related Works} 

In this section, we describe related works in terms of security analysis by machine learning.

\subsection{Security Analysis of Ethereum by Machine Learning}

As machine learning-based analysis, ContractWard~\cite{ContractWard} and the tool by Momeni et al.~\cite{momeni2019machine} are based on support vector machine (SVM) and random forest, whose features are extracted manually by an analyst. 
Hence, identifiable vulnerabilities are limited due to insufficient feature extraction as described in Section~\ref{problem description}. 
As research on automated feature extraction, VulDeeSmartContract~\cite{QL+20} has been proposed by combining Word2Vec with long-short term memory (LSTM), but it specializes only in reentrancy vulnerability. 
Eth2Vec can be considered as a tool dealing with more versatile vulnerabilities in a similar approach. 

Next, EClone~\cite{liu2019enabling} is a tool that detects code clones of Ethereum smart contracts through the computation of code similarity. 
Although EClone does not utilize neural networks, a symbolic analysis tool suitable for vector computations of codes has also been proposed. 
The detectability of Eth2Vec can be improved potentially by combining it with EClone. 

In the neural network-based approach, there is ILF~\cite{he2019learning} which automatically generates an input of fuzzing test via neural networks. 
Loosely speaking, ILF learns codes by extracting features via symbolic execution. 
Since ILF learns outputs by symbolic execution, 
it is an entirely different tool from Eth2Vec, which learns codes themselves. 

Finally, SmartEmbed~\cite{gao2020smartembed} is a tool that identifies bugs of smart contract codes by leveraging FastText, which represents the codes as vectors. 
SmartEmbed is the closest work to Eth2Vec. 
However, SmartEmbed did not discuss an objective function and training algorithm explicitly and just vectorizes codes in accordance with Word2Vec and FastText. 
Consequently, there are several code samples which are unidentified as clones, as shown in the paper~\cite{gao2020smartembed}\footnote{Although the authors of~\cite{gao2020smartembed} describe that the samples are not clones, these sample are type-I\hspace{-.1em}V clones following the definition in ~\cite{Roy07asurvey}.}. 
In contrast, Eth2Vec can precisely extract features of codes with a vast vocabulary and thus can detect vulnerabilities even after codes are rewritten. 
Nevertheless, SmartEmbed is an elegant work that also discussed code repairing as well as versatile bugs. 
Interested readers are advised to read the SmartEmbed paper. 

\subsection{Other Analysis Tools for Ethereum}

To the best of our knowledge, symbolic execution~\cite{luu2016making} is the principal approach for analysis of Ethereum smart contracts. 
Since symbolic execution deals with unknown variables as symbolic variables, 
it is potentially suitable for analysis of smart contracts, which utilizes information outside codes~\cite{chinen2020hunting}, i.e., blockchain. 
Hence, many tools have been proposed so far~\cite{nikolic2018finding,torres2018osiris,liu2018s,chen2017under,feist2019slither,mossberg2019manticore,mueller2018smashing}. 
The primary motivation of recent works aims to extend analysis areas, e.g., inter-contract analysis and contract creation. 
Although symbolic execution consumes a longer time in comparison with a machine learning-based approach as described in Section~\ref{Introduction}, the vulnerability detection of Eth2Vec may be improved by combining it with such tools. 

Another approach for smart contract analysis is formal verification~\cite{bhargavan2016formal} which deduces whether a program satisfies a specification via predicate logic. 
In general, formal verification can provide precise analysis by representing the security in a mathematical way. 
However, the verification itself is a rigid and challenging work. 
Many works just formalized a specification of Ethereum and did not achieve the analysis of vulnerabilities~\cite{hildenbrandt2018kevm,tsankov2018securify,kalra2018zeus,grishchenko2018semantic,grishchenko2018semantic,grischchenko2018ethertrust}. 
As the latest work, Grishchenko et al. presented a formal verification tool named eThor~\cite{schneidewind2020ethor} that can analyze vulnerabilities in a versatile way. 
We consider eThor as the most elegant tool in formal verification, and thus readers interested in formal verification are suggested to read the eThor paper.

Finally, there are several tools on dynamic analysis which execute codes themselves~\cite{rodler2018sereum,ferreira2020aegis}. 
However, using dynamic analysis, an analyst needs to implement and execute attack patterns by him-/herself to prevent vulnerabilities in advance. 
Although a universal attack pattern that can capture various attacks was proposed~\cite{ferreira2020aegis}, an analyst still needs to have knowledge about attacks on smart contracts. 
Therefore, analysts with the ability of utilizing a dynamic analysis are limited, 
and thus static analysis is more reasonable. 


\subsection{Security Applications of Natural Language Processing} 

We briefly describe natural language processing applications to the binary analysis and cybersecurity defense as further related works. 
After Shin et al.~\cite{shin2015recognizing} proposed binary analysis based on neural networks, 
many works for binary analysis based on natural language processing, such as Word2Vec, have been proposed in recent years~\cite{zuo2019neural,asm2vec,duan2020deepbindiff,massarallei2019safe}. 
These works are based on neural networks. 
The closest work to Eth2Vec is Asm2Vec~\cite{asm2vec}, which focuses on assembly language. 
Eth2Vec can be seen as an extension of the implementation of Asm2Vec. 
The source code of Asm2Vec is publicly available and readers are advised to read that paper in conjunction with our work. 
We also note that the same construction can be implemented with other tools described above. 

Likewise, natural language processing is another attractive area in cybersecurity. 
Walk2Friends, which performs a social relation inference attack~\cite{backes2017walk2friends}, and 
DarkEmbed~\cite{tavabi2018darkembed}, which learns low dimensional distributed representations of darkweb/deepweb discussions, are examples of natural language processing. 
Other examples are Log2Vec~\cite{log2vec} and Attack2Vec~\cite{attack2vec}, which learn more generalized cybersecurity information. 
These tools apply natural language processing to cybersecurity areas, while Eth2Vec is an application to Ethereum smart contracts. 

\section{Conclusion} 

In this paper, we proposed Eth2Vec, a static analysis tool based on machine learning that detects vulnerabilities of Ethereum smart contracts. 
The most striking property of Eth2Vec is the automated feature extraction for each contract by leveraging neural networks for natural language processing. 
Consequently, by extracting features implicitly and incorporating lexical semantics between contracts, the vulnerabilities can be detected with 77.0\% precision even after the codes are rewritten. 
Moreover, reentrancy which is one of the most important  vulnerabilities can be detected with 86.6\% precision. 
We also demonstrated that Eth2Vec outperforms the SVM-based method by Momeni et al.~\cite{momeni2019machine} 
in terms of precision, recall, and F1-score. 
We are preparing the release of the implementation of Eth2Vec via GitHub as well. 

We are in the process of improving detection performance, i.e., precision, recall, and F1-score, for vulnerabilities and their underlying code clones. 
In particular, we intend to realize inter-contract analysis whereby multiple contracts affect each other. 
The performance of Eth2Vec will be improved significantly by introducing a function to obtain such information. 
Further studies, which take inter-contract analysis into account, will thus need to be undertaken.







\bibliography{main.bib}

\begin{thebibliography}{10}

\bibitem{atzei2017survey}
N.~Atzei, M.~Bartoletti, and T.~Cimoli.
\newblock A survey of attacks on ethereum smart contracts (sok).
\newblock In {\em Proc. of POST 2017}, volume 10204 of {\em LNCS}, pages
  164--186. Springer, 2017.

\bibitem{backes2017walk2friends}
M.~Backes, M.~Humbert, J.~Pang, and Y.~Zhang.
\newblock Walk2friends: Inferring social links from mobility profiles.
\newblock In {\em Proc. of CCS 2017}, page 1943^^e2^^80^^931957. ACM, 2017.

\bibitem{bhargavan2016formal}
K.~Bhargavan, A.~Delignat-Lavaud, C.~Fournet, A.~Gollamudi, G.~Gonthier,
  N.~Kobeissi, N.~Kulatova, A.~Rastogi, T.~Sibut-Pinote, N.~Swamy, et~al.
\newblock Formal verification of smart contracts: Short paper.
\newblock In {\em Proc. of PLAS 2016}, pages 91--96. ACM, 2016.

\bibitem{brent2018vandal}
L.~Brent, A.~Jurisevic, M.~Kong, E.~Liu, F.~Gauthier, V.~Gramoli, R.~Holz, and
  B.~Scholz.
\newblock Vandal: A scalable security analysis framework for smart contracts.
\newblock {\em arXiv preprint arXiv:1809.03981}, 2018.

\bibitem{chen2017under}
T.~Chen, X.~Li, X.~Luo, and X.~Zhang.
\newblock Under-optimized smart contracts devour your money.
\newblock In {\em Proc. of SANER 2017}, pages 442--446. IEEE, 2017.

\bibitem{chinen2020hunting}
Y.~Chinen, N.~Yanai, J.~P. Cruz, and S.~Okamura.
\newblock Hunting for re-entrancy attacks in ethereum smart contracts via
  static analysis.
\newblock {\em arXiv preprint arXiv:2007.01029}, 2020.

\bibitem{di2019survey}
M.~Di~Angelo and G.~Salzer.
\newblock A survey of tools for analyzing ethereum smart contracts.
\newblock In {\em Proc. of DAPPCON 2019}, pages 69--78. IEEE, 2019.

\bibitem{Ding2016Kam1n0}
S.~H. Ding, B.~C. Fung, and P.~Charland.
\newblock Kam1n0: Mapreduce-based assembly clone search for reverse
  engineering.
\newblock In {\em Proc. of KDD 2016}, page 461^^e2^^80^^93470. ACM, 2016.

\bibitem{asm2vec}
S.~H. Ding, B.~C. Fung, and P.~Charland.
\newblock Asm2vec: Boosting static representation robustness for binary clone
  search against code obfuscation and compiler optimization.
\newblock In {\em Proc. of IEEE S\&P 2019}, pages 472--489. IEEE, 2019.

\bibitem{solidityDoc}
S.~C. S.~. documentation.
\newblock
  https://solidity.readthedocs.io/en/v0.5.11/security-considerations.html.

\bibitem{duan2020deepbindiff}
Y.~Duan, X.~Li, J.~Wang, and H.~Yin.
\newblock Deepbindiff: Learning program-wide code representations for binary
  diffing.
\newblock In {\em Proc. of NDSS 2020}. Internet Society, 2020.

\bibitem{feist2019slither}
J.~Feist, G.~Grieco, and A.~Groce.
\newblock Slither: A static analysis framework for smart contracts.
\newblock In {\em Proc. of WETSEB 2019}, pages 8--15. IEEE, 2019.

\bibitem{ferreira2020aegis}
C.~Ferreira~Torres, M.~Steichen, R.~Norvill, B.~Fiz~Pontiveros, and H.~Jonker.
\newblock {\AE}gis: Shielding vulnerable smart contracts against attacks.
\newblock In {\em Proc. of AsiaCCS 2020}, page 584^^e2^^80^^93597. ACM, 2020.

\bibitem{gao2020smartembed}
Z.~Gao, L.~Jiang, X.~Xia, D.~Lo, and J.~Grundy.
\newblock Checking smart contracts with structural code embedding.
\newblock {\em IEEE Transactions on Software Engineering}, pages 1--1, 2020.

\bibitem{grischchenko2018ethertrust}
I.~Grishchenko, M.~Maffei, and C.~Schneidewind.
\newblock Foundations and tools for the static analysis of ethereum smart
  contracts.
\newblock In {\em Proc. of CAV 2018}, volume 10981 of {\em LNCS}, pages 51--78.
  Springer, 2018.

\bibitem{grishchenko2018semantic}
I.~Grishchenko, M.~Maffei, and C.~Schneidewind.
\newblock A semantic framework for the security analysis of ethereum smart
  contracts.
\newblock In {\em Proc. of POST 2018}, volume 10804 of {\em LNCS}, pages
  243--269. Springer, 2018.

\bibitem{he2019learning}
J.~He, M.~Balunoviundefined, N.~Ambroladze, P.~Tsankov, and M.~Vechev.
\newblock Learning to fuzz from symbolic execution with application to smart
  contracts.
\newblock In {\em Proc. of CCS 2019}, page 531^^e2^^80^^93548. ACM, 2019.

\bibitem{hildenbrandt2018kevm}
E.~Hildenbrandt, M.~Saxena, N.~Rodrigues, X.~Zhu, P.~Daian, D.~Guth, B.~Moore,
  D.~Park, Y.~Zhang, A.~Stefanescu, et~al.
\newblock Kevm: A complete formal semantics of the ethereum virtual machine.
\newblock In {\em Proc. of CSF 2018}, pages 204--217. IEEE, 2018.

\bibitem{hill2016learning}
F.~Hill, K.~Cho, and A.~Korhonen.
\newblock Learning distributed representations of sentences from unlabelled
  data.
\newblock In {\em Proc. of NAACL HLT 2016}, pages 1367--1377. ACL, 2016.

\bibitem{kalra2018zeus}
S.~Kalra, S.~Goel, M.~Dhawan, and S.~Sharma.
\newblock Zeus: Analyzing safety of smart contracts.
\newblock In {\em Proc. of NDSS 2018}. Internet Society, 2018.

\bibitem{le2014distributed}
Q.~Le and T.~Mikolov.
\newblock Distributed representations of sentences and documents.
\newblock In {\em Proc. of ICML 2014}, pages 1188--1196, 2014.

\bibitem{log2vec}
F.~Liu, Y.~Wen, D.~Zhang, X.~Jiang, X.~Xing, and D.~Meng.
\newblock Log2vec: A heterogeneous graph embedding based approach for detecting
  cyber threats within enterprise.
\newblock In {\em Proc. of CCS 2019}. ACM, 2019.

\bibitem{liu2018s}
H.~Liu, C.~Liu, W.~Zhao, Y.~Jiang, and J.~Sun.
\newblock S-gram: towards semantic-aware security auditing for ethereum smart
  contracts.
\newblock In {\em Proc. of ASE 2018}, pages 814--819. ACM, 2018.

\bibitem{liu2019enabling}
H.~Liu, Z.~Yang, Y.~Jiang, W.~Zhao, and J.~Sun.
\newblock Enabling clone detection for ethereum via smart contract birthmarks.
\newblock In {\em Proc. of ICPC 2019}, pages 105--115. IEEE, 2019.

\bibitem{liu2018eclone}
H.~Liu, Z.~Yang, C.~Liu, Y.~Jiang, W.~Zhao, and J.~Sun.
\newblock Eclone: Detect semantic clones in ethereum via symbolic transaction
  sketch.
\newblock In {\em Proc. of ESEC/FSE 2018}, page 900^^e2^^80^^93903. ACM, 2018.

\bibitem{luu2016making}
L.~Luu, D.-H. Chu, H.~Olickel, P.~Saxena, and A.~Hobor.
\newblock Making smart contracts smarter.
\newblock In {\em Proc. of CCS 2016}, pages 254--269. ACM, 2016.

\bibitem{massarallei2019safe}
L.~Massarelli, G.~A. Di~Luna, F.~Petroni, R.~Baldoni, and L.~Querzoni.
\newblock Safe: Self-attentive function embeddings for binary similarity.
\newblock In R.~Perdisci, C.~Maurice, G.~Giacinto, and M.~Almgren, editors,
  {\em Proc. of DIMVA 2019}, volume 11543 of {\em LNCS}, pages 309--329.
  Springer, 2019.

\bibitem{mikolov2013distributed}
T.~Mikolov, I.~Sutskever, K.~Chen, G.~S. Corrado, and J.~Dean.
\newblock Distributed representations of words and phrases and their
  compositionality.
\newblock In {\em Proc. of NIPS 2013}, pages 3111--3119, 2013.

\bibitem{momeni2019machine}
P.~Momeni, Y.~Wang, and R.~Samavi.
\newblock Machine learning model for smart contracts security analysis.
\newblock In {\em Proc. of PST 2019}, pages 1--6. IEEE, 2019.

\bibitem{mossberg2019manticore}
M.~Mossberg, F.~Manzano, E.~Hennenfent, A.~Groce, G.~Grieco, J.~Feist,
  T.~Brunson, and A.~Dinaburg.
\newblock Manticore: A user-friendly symbolic execution framework for binaries
  and smart contracts.
\newblock {\em arXiv preprint arXiv:1907.03890}, 2019.

\bibitem{mueller2018smashing}
B.~Mueller.
\newblock Smashing smart contracts.
\newblock In {\em Proc. of HITBSECCONF 2018}, 2018.

\bibitem{nikolic2018finding}
I.~Nikoli{\'c}, A.~Kolluri, I.~Sergey, P.~Saxena, and A.~Hobor.
\newblock Finding the greedy, prodigal, and suicidal contracts at scale.
\newblock In {\em Proc. of ACSAC 2018}, pages 653--663. ACM, 2018.

\bibitem{norvill2018visual}
R.~Norvill, B.~B.~F. Pontiveros, R.~State, and A.~Cullen.
\newblock Visual emulation for ethereum's virtual machine.
\newblock In {\em Proc of NOMS 2018}, pages 1--4. IEEE, 2018.

\bibitem{QL+20}
P.~Qian, Z.~Liu, Q.~He, R.~Zimmermann, and X.~Wang.
\newblock Towards automated reentrancy detection for smart contracts based on
  sequential models.
\newblock {\em IEEE Access}, 8:19685--19695, 2020.

\bibitem{rodler2018sereum}
M.~Rodler, W.~Li, G.~O. Karame, and L.~Davi.
\newblock Sereum: Protecting existing smart contracts against re-entrancy
  attacks.
\newblock In {\em Proc. of NDSS 2019}. Internet Society, 2019.

\bibitem{Roy07asurvey}
C.~K. Roy and J.~R. Cordy.
\newblock A survey on software clone detection research.
\newblock {\em School of Computing TR 2007-541, Queen's University},
  541(115):64--68, 2007.

\bibitem{schneidewind2020ethor}
C.~Schneidewind, I.~Grishchenko, M.~Scherer, and M.~Maffei.
\newblock Ethor: Practical and provably sound static analysis of ethereum smart
  contracts.
\newblock In {\em Proc. of CCS 2020}, page 621^^e2^^80^^93640. ACM, 2020.

\bibitem{attack2vec}
Y.~Shen and G.~Stringhini.
\newblock Attack2vec: Leveraging temporal word embeddings to understand the
  evolution of cyberattacks.
\newblock In {\em Proc. of {USENIX} Security 2019}, pages 905--921. {USENIX}
  Association, 2019.

\bibitem{shin2015recognizing}
E.~C.~R. Shin, D.~Song, and R.~Moazzezi.
\newblock Recognizing functions in binaries with neural networks.
\newblock In {\em Proc. of {USENIX} Security 2015}, pages 611--626. {USENIX}
  Association, 2015.

\bibitem{song2019efficient}
J.~Song, H.~He, Z.~Lv, C.~Su, G.~Xu, and W.~Wang.
\newblock An efficient vulnerability detection model for ethereum smart
  contracts.
\newblock In {\em Proc. of NSS 2019}, volume 11928 of {\em LNCS}, pages
  433--442. Springer, 2019.

\bibitem{suiche2017porosity}
M.~Suiche.
\newblock Porosity: A decompiler for blockchain-based smart contracts bytecode.
\newblock {\em Proc. of DEFCON 2017}, 25:11, 2017.

\bibitem{tavabi2018darkembed}
N.~Tavabi, P.~Goyal, M.~Almukaynizi, P.~Shakarian, and K.~Lerman.
\newblock Darkembed: Exploit prediction with neural language models.
\newblock In {\em Proc. of AAAI 2018}, pages 7849--7854. {AAAI} Press, 2018.

\bibitem{tikhomirov2018smartcheck}
S.~Tikhomirov, E.~Voskresenskaya, I.~Ivanitskiy, R.~Takhaviev, E.~Marchenko,
  and Y.~Alexandrov.
\newblock Smartcheck: Static analysis of ethereum smart contracts.
\newblock In {\em Proc. of WETSEB 2018}, pages 9--16. ACM, 2018.

\bibitem{torres2018osiris}
C.~F. Torres, J.~Sch{\"u}tte, et~al.
\newblock Osiris: Hunting for integer bugs in ethereum smart contracts.
\newblock In {\em Proc. of ACSAC 2018}, pages 664--676. ACM, 2018.

\bibitem{tsankov2018securify}
P.~Tsankov, A.~Dan, D.~Drachsler-Cohen, A.~Gervais, F.~Buenzli, and M.~Vechev.
\newblock Securify: Practical security analysis of smart contracts.
\newblock In {\em Proc. of CCS 2018}, pages 67--82. ACM, 2018.

\bibitem{ContractWard}
W.~Wang, J.~Song, G.~Xu, Y.~Li, H.~Wang, and C.~Su.
\newblock Contractward: Automated vulnerability detection models for ethereum
  smart contracts.
\newblock {\em IEEE Transactions on Network Science and Engineering}, pages
  1--1 (Early Access), 2020.

\bibitem{Weiss2019annotary}
K.~Weiss and J.~Sch{\"{u}}tte.
\newblock {Annotary: A Concolic Execution System for Developing Secure Smart
  Contracts}.
\newblock In {\em Proc. of ESORICS 2019}, volume 11735 of {\em LNCS}, pages
  747--766. Springer, 2019.

\bibitem{theyellowpaper}
G.~Wood.
\newblock Ethereum: A secure decentralised generalised transaction ledger
  byzantium version.
\newblock https://ethereum.github.io/yellowpaper/paper.pdf.

\bibitem{zhou2018erays}
Y.~Zhou, D.~Kumar, S.~Bakshi, J.~Mason, A.~Miller, and M.~Bailey.
\newblock Erays: reverse engineering ethereum's opaque smart contracts.
\newblock In {\em Proc. of USENIX Security 2018}, pages 1371--1385. Usenix
  Association, 2018.

\bibitem{zou2019smart}
W.~Zou, D.~Lo, P.~S. Kochhar, X.-B.~D. Le, X.~Xia, Y.~Feng, Z.~Chen, and B.~Xu.
\newblock Smart contract development: Challenges and opportunities.
\newblock {\em IEEE Transactions on Software Engineering}, pages 1--1, 2019.

\bibitem{zuo2019neural}
F.~Zuo, X.~Li, P.~Young, L.~Luo, Q.~Zeng, and Z.~Zhang.
\newblock Neural machine translation inspired binary code similarity comparison
  beyond function pairs.
\newblock In {\em Proc. of NDSS 2019}. Internet Society, 2019.

\end{thebibliography}
\bibliographystyle{abbrv}

\appendix

\subsection{Output Example by Eth2Vec} \label{output example}

\begin{figure}[ht]
  \centering
  \includegraphics[width=\columnwidth]{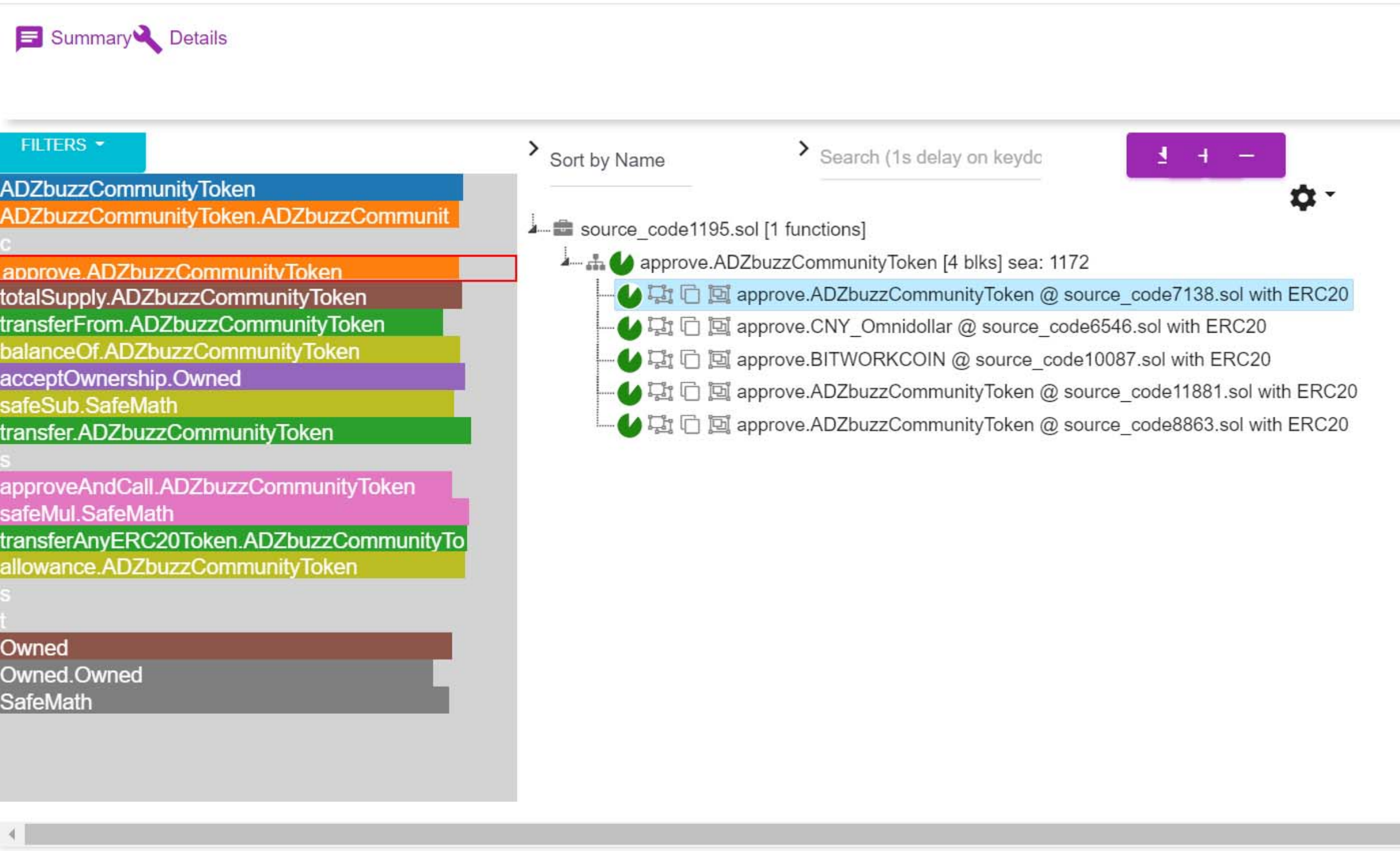} 
  \caption{Output example by Eth2Vec: The left-side on the windows displays a list of functions contained in each contract and the right-side shows a list of vulnerabilities included in the chosen function. 
  }
  \label{fig:output}
\end{figure}

We show an example of an output by Eth2Vec in Fig.~\ref{fig:output}. 
The interface of Eth2Vec follows that of \texttt{Kam1n0}~\cite{Ding2016Kam1n0} as described in Section~\ref{implementation}. 

\subsection{Feature Selection by Momeni et al.} \label{feature by Momeni}

Momeni et al.~\cite{momeni2019machine} extracted 16 features from an abstract syntax tree (AST) of an Ethereum smart contract. 
Among them, we did not adopt Hexadecimal addresses as described in Section~\ref{experiments}. 
The whole list of vulnerabilities as follows, where all the information can be obtained from AST except for hexadecimal addresses: 
\begin{itemize}
    \item Lines of codes
    \item Contract definitions
    \item Function definitions
    \item Binary operations
    \item Function calls
    \item Blocks
    \item Expression statements
    \item Event definitions
    \item Bytes
    \item Elementary type addresses
    \item Modifier definitions
    \item Placeholder statements
    \item Modifier invocation
    \item Approve function definitions
    \item Constant values
    \item Hexadecimal addresses 
\end{itemize}

\subsection{List of Vulnerabilities} \label{vulnerability list}

\begin{table}[ht]
  \begin{center}
      \caption{List of vulnerabilities: In this paper, we target the following vulnerabilities for evaluation of Eth2Vec.}
    \label{tab:vul_list}
    \begin{tabular}{|l|c|l|}
      \hline
     Name & Severity & Description \\  \hline \hline
      \multirow{2}{*}{Reentrancy} & \multirow{2}{*}{3} & 
      External contracts should be called \\
      & & after all local state updates \\ \hline 
      \multirow{2}{*}{Time Dependency} & \multirow{2}{*}{2} & 
      Miners can alter timestamps. Make critical \\
      & & code independent of the environment \\ \hline 
      \multirow{2}{*}{ERC-20 Transfer} & \multirow{2}{*}{1} & 
      The contract throws where the ERC20 standard \\
      & & expects a bool. 
      Return \texttt{false} instead \\ \hline 
      \multirow{2}{*}{Gas Consumption} & \multirow{2}{*}{1} & 
      A transaction fails by exceeding an upper \\
      & & bound on the amount of gas that can be spent \\ \hline 
      \multirow{2}{*}{Implicit Visibility} & \multirow{2}{*}{1} & 
      Functions are public by default. Avoid \\
      & & ambiguity: explicitly declare visibility level \\ \hline 
      Integer Overflow & \multirow{2}{*}{1} & 
      The return value is not checked. \\
      Integer Underflow & & 
      Always check return values of functions \\ \hline 
    \end{tabular}
  \end{center}
\end{table}

In this paper, we target the vulnerabilities described in Table~\ref{tab:vul_list} for evaluation of Eth2Vec. 
The columns of Severity and Description follow descriptions in SmartCheck~\cite{tikhomirov2018smartcheck} except for gas consumption. Description about the gas consumption follows description in \url{https://consensys.github.io/smart-contract-best-practices/known_attacks/}.


\end{document}